\documentclass[pre,showpacs,twocolumn,floats,epsf]{revtex4}
\usepackage{graphicx}
\usepackage{dcolumn}
\usepackage{epsf}
\usepackage{amsmath}

\DeclareSymbolFont{AMSb}{U}{msb}{m}{n}
\DeclareMathSymbol{\N}{\mathbin}{AMSb}{"4E}
\DeclareMathSymbol{\Z}{\mathbin}{AMSb}{"5A}
\DeclareMathSymbol{\R}{\mathbin}{AMSb}{"52}
\DeclareMathSymbol{\Q}{\mathbin}{AMSb}{"51}
\DeclareMathSymbol{\I}{\mathbin}{AMSb}{"49}
\DeclareMathSymbol{\C}{\mathbin}{AMSb}{"43}

\def\openone{\leavevmode\hbox{\small1\kern-3.3pt\normalsize1}}
\newcommand{\la}{\langle}
\newcommand{\ra}{\rangle}

\newcommand{\be}{\begin{equation}}
\newcommand{\ee}{\end{equation}}
\newcommand{\bea}{\begin{eqnarray}}
\newcommand{\eea}{\end{eqnarray}}

\begin{document}

\title{Nonequilibrium work distribution of a quantum harmonic oscillator} 
\author{Sebastian Deffner and Eric Lutz \\
{\it {\small Institute of Physics, University of Augsburg, D-86135 Augsburg, Germany}}}
\date{\today}

\begin{abstract}
We analytically calculate the work distribution of a quantum harmonic oscillator with arbitrary time--dependent angular frequency. We provide detailed expressions for the work probability density for adiabatic and nonadiabatic processes, in the limit of low and high temperature. We further verify the validity of the quantum Jarzynski equality.
\end{abstract}
\pacs{05.30.-d, 05.70.Ln, 05.40.-a}
\maketitle

\section{Introduction}
Fluctuations play  a fundamental role in small nonequilibrium systems. While thermodynamic quantities like the free energy difference can easily be calculated for slow, quasistatic processes from the knowledge of the initial and final equilibrium partition functions, $\Delta F= F_1-F_0= -kT \ln Z_1/Z_0$, a general method for evaluating $\Delta F $ in the case of fast, nonequilibrium transformations was lacking until recently. A decade ago, Jarzynski introduced a remarkable equality that relates the free energy difference to the averaged exponentiated work done during the process \cite{jar97},
\be
\label{eq1}
\Delta F = -kT \ln \la e^{-\beta W}\ra \ ,
\ee
with $\la e^{-\beta W}\ra = \int dW e^{-\beta W} \,P(W)$ and $\beta=1/kT$ the inverse temperature. Equation (\ref{eq1}) allows to compute the difference $\Delta F$ for arbitrary transformations, quasistatic or not, once the corresponding distribution of work $P(W)$ is known. The free energy difference is thus entirely determined by the fluctuations of the total work $W$, a property which emphasizes the importance of the work probability density. The usefulness of Eq.~(\ref{eq1}) for evaluating $\Delta F$ in single biomolecules has  been demonstrated in DNA unfolding experiments \cite{lip02,bus05}. 

Extensions of the classical Jarzynski equality (\ref{eq1}) to quantum--mechanical systems have been discussed in Refs.~\cite{tas00,muk03,mae04,che04,mon05,tal07a,tal07b}. In that respect, it is worth mentioning that  the concept of work in  quantum thermodynamics has been  clarified in Ref.~\cite{tal07a}. Thus for a quantum Hamiltonian system, work cannot be considered as the expectation value of some work operator, in  contrast to some other thermodynamic quantities like  energy. Rather the exponential average of the work, as well as its characteristic function, are given by time-ordered quantum correlation functions.

Our aim in this paper is to determine the work distribution for a simple quantum system, namely a time--dependent harmonic oscillator. We obtain the exact analytical expression of the characteristic function of the work and provide approximations for the work distribution for adiabatic and  nonadiabatic processes, in the limit of low and high temperature. Moreover, we check the validity of the Jarzynski equality for arbitrary processes.

\section{Work characteristic function}
We consider a quantum--mechanical harmonic oscillator with a time-dependent angular frequency $\omega(t)$. The corresponding  Hamiltonian is of the usual form
\begin{equation}
\label{eq2}
H=\frac{p^2}{2m}+\frac{m}{2}\omega^2(t)x^2 \ .
\end{equation}
We shall be interested in calculating the work distribution of the oscillator when the angular frequency is changed from an initial value  
$\omega_0$ at $t=0$ to a final value $\omega_1$ at $t= \tau$. 
We denote by $\phi_n^t$ the eigenfunctions and by $E_n^t= \hbar\omega(t)(n+\frac{1}{2})$  the eigenvalues  of the Hamiltonian (\ref{eq2}) at any given time $t$. We make the additional assumption that the oscillator is initially thermalized with inverse temperature $\beta$ but is  otherwise isolated. 

The probability density of the total work done on the harmonic oscillator during time $\tau$ is given by \cite{tal07a}
\begin{equation}
\label{eq4}
P(W)=\sum\limits_{m,n}\,\delta[W-(E_{m}^{\tau }-E_{n}^{0})]\,P_{m,n}^{\tau}\, P^0_{n} \ ,
\end{equation}
where $P_n^0=\exp(-\beta E_n^0)/Z_0$ is the initial (thermal) occupation probability and $P_{m,n}^{\tau}$ are the transition probabilities between initial and final states $n$ and $m$, 
\begin{equation}
\label{eq5}
P_{m,n}^{\tau}=\Big|\int dx_0\int dx \, {\phi_m^\tau}^*(x)U(x,x_0;\tau)\phi^0_n(x_0)\Big|^2 \ .
\end{equation}
The evaluation of the transition probabilities (\ref{eq5}) requires the computation of the propagator $ U(x,x_0;\tau)$ of the time--dependent harmonic oscillator. To this end, we  follow the approach developed by Husimi  \cite{2}.

The dynamics generated by  the quadratic Hamiltonian (\ref{eq2}) is Gaussian for any $\omega(t)$. By introducing the  Gaussian wave function ansatz,
\begin{equation}
\label{eq6}
\psi(x,t)=e^{\frac{i}{2\hbar}\big(a(t)x^2+2b(t)x+c(t)\big)}\ , 
\end{equation}
the  Schr\"odinger equation for the quantum oscillator can be reduced to  a system of three coupled differential equations for the parameters $a, b$ and $c$,
\begin{eqnarray}
\frac{1}{m}\frac{da}{dt}&=&-\frac{a^2}{m^2}-\omega^2(t)\ , \label{14}\\
\frac{db}{dt}&=&-\frac{a}{m}b \label{15}\ ,\\ 
\frac{dc}{dt}&=&i\hbar\frac{a}{m}-\frac{1}{m}b^2 \ .\label{16}
\end{eqnarray}
The nonlinear equation (\ref{14}) is of the Riccati type. Writing $a=\frac{m\dot X}{X}$, it can be mapped to the equation of motion of a classical time--dependent linear oscillator,
\begin{equation}
\frac{d^2X}{dt^2}+\omega^2(t)X=0 \ .\label{eq9}
\end{equation}
The solutions of Eqs.~(\ref{14})--(\ref{16}), and hence the wave function (\ref{eq6}) of the quantum oscillator, can be recurrently constructed from the solution of Eq.~(\ref{eq9}). An explicit solution of 
Eq.~(\ref{eq9}) for the linear parametrization \cite{jar97a},
 \begin{equation}
\label{eq3}
\omega^2(t)=\omega_0^2-(\omega_0^2-\omega_1^2)\frac{t}{\tau} \ ,
\end{equation}
is given in Appendix A.
 We stress  that the following results are valid for arbitrary $\omega(t)$. 

The general form of the propagator can be determined from the wave function (\ref{eq6}) by noting that $\phi(x,\tau) = \int dx \,U(x,x_0;\tau)\, \phi(x_0,0)$. It is explicitly given by
\be
\label{eq12}
U(x,x_0;\tau) = \sqrt{\frac{m}{2\pi i h X}} \exp\Big[\frac{i m}{2\hbar X} (\dot Xx^2-2 xx_0+ Y x_0^2) \Big] \ ,
\ee
where $X(t)$ and $Y(t)$ are solutions of Eq.~(\ref{eq9}) satisfying
\begin{eqnarray}
\label{eq12a}
X(0)=0 & \dot X(0)=1  \nonumber\ , \label{2}\\
Y(0)=1 & \dot Y(0)=0  \label{18}\ .
\end{eqnarray}
A compact expression for the generating function of the transition probabilities $P_{m,n}^{\tau}$ can then  be obtained by using the generating functions of the eigenfunctions $\phi_n^t(x)$ of the quantum harmonic oscillator, 

\begin{eqnarray}
\label{eq13}
&\sum\limits_{n=0}^\infty\,u^n\phi_n^t(x){\phi^t_n}^*(x_0) =\nonumber\\
&\sqrt{\frac{m\omega(t)}{\hbar\pi^2(1-u^2)}}\exp\Big[-\frac{m\omega(t)}{\hbar}\frac{(1+u^2)(x^2+x_0^2)-4u xx_0}{2(1-u^2)}\Big] \ .
\end{eqnarray}
By exploiting the Gaussian character of Eqs.~(\ref{eq12}) and (\ref{eq13}), the generating function of Eq.~(\ref{eq5}) becomes
\begin{eqnarray}
\label{eq14}
P(u,v)=&\sum\limits_{m,n}\,u^m v^n P_{m,n}^\tau\nonumber\\ 
=& \frac{\sqrt2}{\sqrt{Q^{*}(1-u^2)(1-v^2)+(1+u^2)(1+v^2)-4uv}} \ ,
\end{eqnarray}
where we have introduced the quantity
\begin{equation}
\label{eq15}
Q^{*}=\frac{1}{2\omega_0\omega_1}\big[\omega_0^2\,(\omega_1^2\,X(\tau)^2+\dot X(\tau)^2)+(\omega_1^2\,Y(\tau)^2+\dot Y(\tau)^2)\big] 
\end{equation}
The parameter $Q^*$ will play an important role in the following discussion. We will indeed see that it can be regarded as a measure of the degree of adiabaticity of the process. An interesting feature of the generating function $P(u,v)$ is further that its $(u,v)$--dependence  remains the same for all possible transformations $\omega(t)$. Details about the specific parametrization of the angular frequency only enter through different numerical values of $Q^*$.

With the help of Eq.~(\ref{eq14}), the characteristic function of the work, defined as the Fourier transform of the  probability distribution (\ref{eq4}),  
\begin{eqnarray}
G(\mu) &=& \int dW e^{i \mu W} \,P(W)\nonumber \\
&=&\frac{e^{-\frac{\beta}{2}\,\hbar \omega_{0}}}{Z_{0}}\, e^{-i \mu \frac{\hbar \omega_{0}}{2}} e^{i \mu \frac{\hbar \omega_{\tau}}{2}}\nonumber \\
&\times&\sum_{m,n} [e^{i \mu \hbar \omega_{1}}]^m [e^{-(i \mu +\beta)\hbar \omega_0}]^n \,P_{m,n}^{\tau} \ ,
\end{eqnarray}
can eventually be written in closed form in terms of the energies, $\varepsilon_0=\hbar\omega_0$, $\varepsilon_1=\hbar\omega_1$ ($\Delta\varepsilon=\varepsilon_1-\varepsilon_0$) and the inverse temperature $\beta$. We have
\begin{widetext}
\begin{eqnarray}
\label{eq17}
G(\mu)&=&\frac{\sqrt{2}(1-e^{-\beta\varepsilon_0})\, e^{i\mu \frac{\Delta\varepsilon}{2}}}{\sqrt{Q^{*}(1-e^{2i \mu \varepsilon_1})(1-e^{-2(i \mu +\beta)\varepsilon_0})+(1+e^{2i \mu \varepsilon_1})(1+e^{-2(i \mu +\beta)\varepsilon_0})-4\,e^{i \mu \varepsilon_1} e^{-(i \mu +\beta)\varepsilon_0}}} \ .
\end{eqnarray}
\end{widetext}
The above expression for $G(\mu)$ is exact and fully characterizes the work distribution of the time--dependent quantum harmonic oscillator (\ref{eq2}), for arbitrary parametrization of the angular frequency $\omega(t)$. As mentioned previously, different realizations of $\omega(t)$ will merely lead to different values of the parameter $Q^*$. One should also note that the mass $m$ of the oscillator does not explicitly appear in (\ref{eq17}).

\subparagraph{Recovering the Jarzynski equality}
As shown in Ref.~\cite{tal07a}, the Jarzynski equality (\ref{eq1}) can be recovered from the characteristic function $G(\mu)$, without having to  compute its inverse Fourier transform, by simply putting $\mu =i\beta$. In this way, we obtain from Eq.~(\ref{eq17}),
\bea
\label{eq18}
G(i \beta)= \langle e^{-\beta W}\rangle 
=\frac{\sinh{\frac{\beta}{2}\varepsilon_0}}{\sinh{\frac{\beta}{2}\varepsilon_1}} = e^{-\beta\Delta F} \ .
\eea
The latter agrees with the direct derivation of the free energy difference of the quantum harmonic oscillator (see for instance Ref.~\cite{cal84}).

\section{Work probability distribution}
The direct analytic evaluation of the nonequilibrium work distribution $P(W)$ by Fourier inverting Eq.~(\ref{eq17}) does not seem to be feasible in the general case. In this section, we derive approximate expressions of the work probability density for adiabatic and  nonadiabatic processes, in the zero and high--temperature limit.
\subparagraph{Adiabatic case}
We base our discussion of adiabaticity on the equivalent classical harmonic oscillator (\ref{eq9}) since the characteristic function $G(\mu)$ is fully determined through its solutions $X(t)$ and $Y(t)$ \cite{2}. For an adiabatic transformation, the action of the oscillator, given by the ratio of the energy to the angular frequency, is a time--independent constant. In other words, for a quasistatic process, we have the two adiabatic invariants,
\begin{equation}
\frac{\dot X^2 + \omega^2(t)X^2}{\omega(t)}=\frac{\dot Y^2 + \omega^2(t)Y^2}{\omega(t)}=\frac{1}{\omega_0} \ .\label{5}
\end{equation}
From the definition (\ref{eq15}) of the parameter $Q^*$, we see that in this case we simply have  $Q^*=1$. As a consequence $P(u,v)=(1-uv)^{-1}$ and $P^\tau_{m,n}= \delta_{m,n}$. The latter  is an expression of the quantum adiabatic theorem: For  infinitely slow transformations, there are no transitions between different quantum states. The characteristic  function (\ref{eq17}) accordingly reduces to 
\begin{equation}
G(\mu)=\frac{(1-e^{-\beta\varepsilon_0})e^{i \mu \frac{\Delta\varepsilon}{2}}}{1-e^{-\beta\varepsilon_0}e^{i\mu\Delta\varepsilon}} \ . \label{6}
\end{equation}
Equation (\ref{6})  further simplifies in the limit of low and high temperature. In the classical limit, $\hbar\beta\ll 1$, we find
\begin{equation}
\label{eq21}
G(\mu)=\frac{\beta\omega_0}{\beta\omega_0-i\Delta\omega\mu} \ .
\end{equation}
The inverse Fourier transform of Eq.~(\ref{eq21}) then yields  the adiabatic  work distribution
\begin{equation}
\label{eq22}
P(W)=\frac{\beta\omega_0e^{-\beta\frac{\omega_0}{\Delta\omega}W}\Theta(W)}{\Delta\omega} \ .
\end{equation}
This result is identical to the  classical work probability distribution derived by  Jarzynski \cite{jar97a}. 

In the opposite limit of low temperature, $\hbar\beta\gg 1$, we have  
\begin{equation}
\label{eq23}
G(\mu)=e^{i\mu\frac{\Delta\varepsilon}{2}}\ .
\end{equation}
As a result, the zero--temperature adiabatic work distribution is given by a delta function,
\begin{equation}
\label{eq24}
P(W)=\delta\Big(W-\frac{\Delta\varepsilon}{2}\Big) \ . 
\end{equation}

\begin{figure} [b]
\epsfxsize=0.46\textwidth
\epsffile{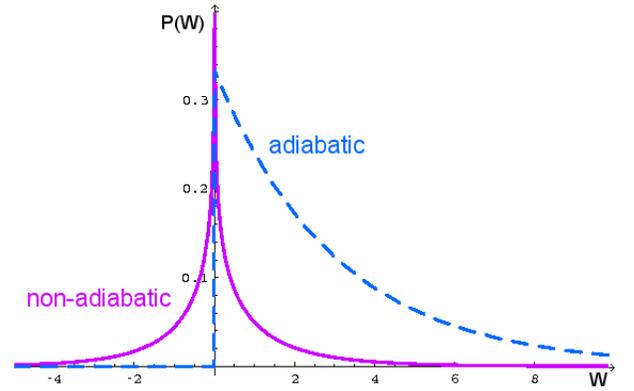}
\caption{Work probability distribution in the classical limit in the adiabatic and nonadiabatic case, Eqs.~(\ref{eq22}) and (\ref{eq27}),  for the parameters $\omega_0=1$, $ \omega_1=1.3$,  $\beta=0.1$ and $Q^*=5$.}
\label{fig}
\end{figure}
\subparagraph{Nonadiabatic case}
We now turn to the nonadiabatic case $Q^* >1$. We start from  the general expression (\ref{eq17}) for the characteristic function and expand it  in the classical limit, $\hbar\beta\ll 1$. We obtain
\be
\label{eq25}
G(\mu) =  \frac{\beta\omega_0}{\sqrt{A\mu^2+B\mu+C}} \ ,
\ee
where we have defined the constants $A=2Q^*\omega_0\omega_1-\omega_0^2-\omega_1^2$,\,\,$B=2i(\beta\omega_0^2-Q^*\beta\omega_0\omega_1)$ and $C=\beta^2\omega_0^2$. In order to perform  the inverse Fourier transform of Eq.~(\ref{eq25}), we use the following integral  \cite{5},
\begin{equation}
\int\limits_{-\infty}^{+\infty}\,dx\frac{e^{i\lambda x}}{(x^2+z^2)^\rho}=2^{\frac{3}{2}-\rho}\sqrt{\pi}\frac{1}{\Gamma(\rho)}\bigg(\frac{|\lambda|}{z}\bigg)^{\rho-\frac{1}{2}}K_{\rho-\frac{1}{2}}(|\lambda| z) \label{8}
\end{equation}
where $\Gamma(x)$ denotes the Euler Gamma function and $K_{\nu}(x)$ is the Macdonald function, i.e the modified Bessel function of the third kind. Combining Eqs.~(\ref{eq25}) and (\ref{8}), we then arrive at the nonadiabatic work distribution, 
\begin{widetext}
\be
\label{eq27}
P(W)=\sqrt{\frac{2 \beta^2\omega_0^2/\pi}{2 Q^*\omega_0\omega_1-\omega_0^2-\omega_1^2}}e^{\Big(\frac{\beta\omega_0^2-Q^*\beta\omega_0\omega_1}{2 Q^*\omega_0\omega_1-\omega_0^2-\omega_1^2}W\Big)} K_0\Bigg[\sqrt{\frac{\beta^2\omega_0^2}{2 Q^*\omega_0\omega_1-\omega_0^2-\omega_1^2}-\frac{(\beta^2\omega_0^2-Q^*\beta\omega_0\omega_1)^2}{(2 Q^*\omega_0\omega_1-\omega_0^2-\omega_1^2)^2}}|W|\Bigg]  \ .
\ee
\end{widetext}

In the low temperature limit, \mbox{$\hbar\beta\gg 1$}, the characteristic function (\ref{eq17}) reads,
\begin{equation}
\label{eq28}
G(\mu)=\frac{\sqrt{2} e^{i\mu\frac{\Delta \varepsilon}{2}}}{\sqrt{Q^*+1-(Q^*-1) e^{2i\mu\varepsilon_1}}} \ .
\end{equation}
The inverse Fourier transform of Eq.~({\ref{eq28}) can be approximated in the limit of small $\varepsilon_1$ by, 
\begin{equation}
\label{eq29}
P(W)=\frac{e^{-\frac{W-\Delta \varepsilon/2}{(Q^*-1)\varepsilon1}}}{\sqrt{\pi (Q^*-1) \varepsilon1(W-\Delta \varepsilon/2)}}\ .
\end{equation}
The zero--temperature nonadiabatic work distribution (\ref{eq29}) is valid when $W\geq\frac{\Delta\varepsilon}{2}$.

\section{Discussion}
The parameter $Q^*$ defined in Eq.~(\ref{eq15}) can be ascribed a simple physical meaning \cite{2}. Let us  consider a transition from initial state $n$ to final state $m$. Then the average quantum number in the final state is
\be
\label{eq30}
\la m\ra_n = \sum_m m P_{m,n}^\tau = (n+\frac{1}{2}) Q^* - \frac{1}{2} \ ,
\ee 
which follows from the expression of the generating function (\ref{eq14}) of  $P_{m,n}^\tau $.
Similarly, the mean--square quantum number after time $\tau$ is given by 
\be
\label{eq31}
\sigma^2_{m,n} = \la \left (m-\la m \ra_n\right)^2\ra_n = \frac{1}{2} ({Q^*}^2 -1)(n^2+n+1) \ .
\ee
Equation (\ref{eq31}) shows that  the parameter $Q^*$ directly controls the magnitude of the variance $\sigma^2_{m,n}$. In the adiabatic limit, $Q^*=1$, we readily  get $\la m\ra_n = n $ and $\sigma^2_{m,n}=0$. We therefore recover that for quasistatic  processes, the  system remains in its initial state, $m=n$. On the other hand, for fast nonadiabatic processes, the mean $\la m\ra_n$ and the dispersion $\sigma^2_{m,n}$ increase with increasing values of $Q^*$, indicating that the quantum oscillator ends in a final state $m$  which is farther and farther away from the initial state $n$. The behavior of $Q^*$ as a function of the inverse switching time $1/\tau$ is shown in Fig.~(3) for the linear parametrization (\ref{eq3}) with fixed values of the initial and final angular frequencies. We observe, as expected, that infinitely slow switching, $1/\tau \rightarrow 0$, corresponds to the adiabatic limit $Q^*\rightarrow 1$, while, for finite--time transformations, the parameter  $Q^*$   monotonically increases  with   the inverse switching time.  
\begin{figure}[t]
\epsfxsize=.97\columnwidth \epsffile{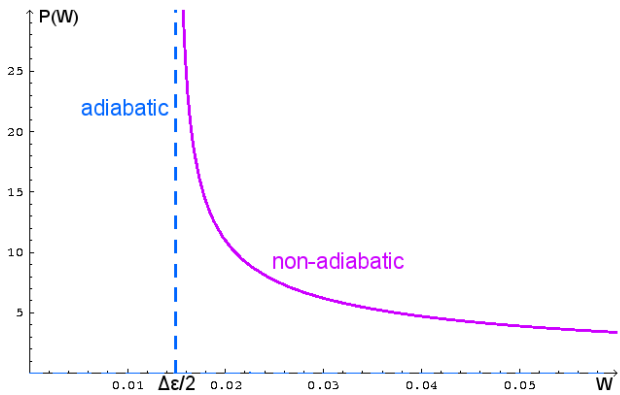}
\caption {Work probability distribution at zero temperature  in the adiabatic and nonadiabatic case,  Eqs.~(\ref{eq24}) and (\ref{eq29}),  for the parameters $\varepsilon_0=0.1$, $\varepsilon_1=0.13$ and $Q^*=5$.}
\end{figure}

The adiabatic  and nonadiabatic  work distributions Eqs.~(\ref{eq22}), ({\ref{eq24}) and Eqs.~({\ref{eq27}), (\ref{eq29}) are plotted in Figs.~(1) and (2). They illustrate the effect of temperature and nonadiabaticity on the probability density $P(W)$ of the harmonic oscillator. The mean and the variance of the work probability distribution, as well as the mean dissipated work can be found in Appendix B. Let us begin with Fig.~(2).  For an adiabatic transformation at zero temperature, the oscillator remains  in the ground state for all times. As a consequence, the total work and the free energy difference are  equal to the energy difference between the initial and final ground states, $W= \Delta F = \Delta \varepsilon/2$. In this case,  $P(W)$ is simply a delta function. As we now increase the temperature, the initial thermal distribution broadens. This then results in a widening  of the work distribution to a one--sided exponential in the high--temperature limit as seen in Fig.~(1). On the other hand, when the nonadiabaticity parameter $Q^*$ is augmented, we observe still in Fig.~(1) that the work distribution acquires a negative tail. This is a remarkable fact since for a process with  negative total work, the quantum harmonic oscillator is  cooled when its angular frequency is varied and not heated. This situation can obviously not occur when the oscillator is already in its ground state, explaining the absence of a negative tail in the nonadiabatic work distribution in Fig.~(2). Here the broadening of the distribution $P(W)$ is related to the work which is dissipated when the oscillator is brought to an excited state.
\begin{figure}[h]
\epsfxsize=.95\columnwidth \epsffile{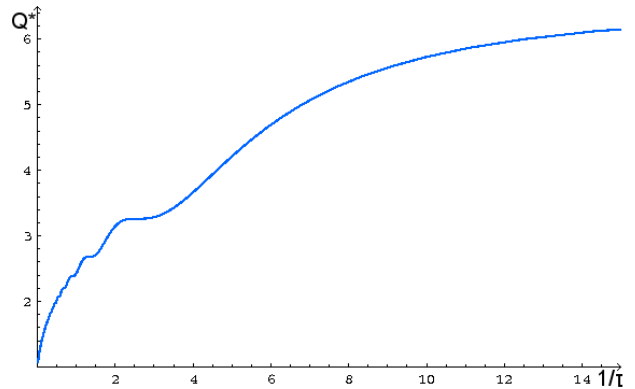}
\caption {Parameter $Q^*$, given by Eq.~(\ref{eq15}), as a function of the inverse switching time $1/\tau$. $Q^*$ is here calculated for the linear parametrization  (\ref{eq3}) with parameters $\omega_0=1$ and $\omega_1= 13$.}
\end{figure}

\section{Summary}
We have analytically calculated the characteristic function of the total work of a quantum harmonic oscillator with a time--dependent angular frequency. We have obtained approximate formulas for the work distribution for the case of adiabatic and nonadiabatic transformations, in the limit of low and high temperature. We have further checked the validity of the quantum Jarzynski equality for arbitrary parametrizations of the angular frequency and discussed the effect of temperature and nonadiabaticity of the work probability density.
\subparagraph{Acknowledgements}
SD and EL would like to thank  Ramses van Zon, Ferdinand Schmidt--Kaler and  Peter H\"anggi for interesting discussions and comments. This work was supported by  the
Emmy Noether Program of the DFG (contract LU1382/1-1) and the
cluster of excellence Nanosystems Initiative Munich (NIM).
\section*{Appendix A}
The equation of motion for the time--dependent harmonic oscillator (\ref{eq9}) can in general not be solved exactly. In this appendix, we provide the analytical solutions of Eq.~(\ref{eq9}) for the linear parametrization (\ref{eq3}) of the angular frequency, obeying the boundary conditions  ({\ref{eq12a}). They can expressed with the help of the  Airy-A and Airy-B functions \cite{17},
\begin{widetext}
\begin{eqnarray}
X(t) & = & \frac{\pi \tau^{\frac{1}{3}}}{[(\omega_0-\omega_1)(\omega_0+\omega_1)]^{\frac{1}{3}}}\Bigg( -Ai\bigg(\frac{(t-\tau)\omega_0^2-t\omega_1^2}{\tau^{\frac{1}{3}}[(\omega_0-\omega_1)(\omega_0+\omega_1)]^{\frac{2}{3}} }\bigg)Bi\bigg(-\frac{\omega_0^2 \tau^{\frac{2}{3}}}{[(\omega_0-\omega_1)(\omega_0+\omega_1)]^{\frac{2}{3}}}\bigg) \nonumber\\
 & + & Ai\bigg(-\frac{\omega_0^2 \tau^{\frac{2}{3}}} {[(\omega_0-\omega_1)(\omega_0+\omega_1)]^{\frac{2}{3}}} \bigg) Bi\bigg(\frac{(t-\tau)\omega_0^2-t\omega_1^2}{\tau^{\frac{1}{3}}[(\omega_0-\omega_1)(\omega_0+\omega_1)]^{\frac{2}{3}}}\bigg)\Bigg) \ ,
\end{eqnarray}
and
\begin{eqnarray}
Y(t)&=&\pi\Bigg(Ai'\bigg(-\frac{\omega_0^2 \tau^{\frac{2}{3}}} {[(\omega_0-\omega_1)(\omega_0+\omega_1)]^{\frac{2}{3}}} \bigg) Bi\bigg(\frac{(t-\tau)\omega_0^2-t\omega_1^2}{\tau^{\frac{1}{3}}[(\omega_0-\omega_1)(\omega_0+\omega_1)]^{\frac{2}{3}}}\bigg)\Bigg)\nonumber\\
 &+& Ai\bigg(\frac{(t-\tau)\omega_0^2-t\omega_1^2}{\tau^{\frac{1}{3}}[(\omega_0-\omega_1)(\omega_0+\omega_1)]^{\frac{2}{3}}}\bigg)\Bigg)Bi'\bigg(-\frac{\omega_0^2 \tau^{\frac{2}{3}}} {[(\omega_0-\omega_1)(\omega_0+\omega_1)]^{\frac{2}{3}}} \bigg)\Bigg) \ , 
\end{eqnarray}
\end{widetext}
where $Ai'(x)$ and $Bi'(x)$ denote the first derivatives with respect to $x$.

\section*{Appendix B}
In this appendix, we evaluate the mean and the variance of the work probability distribution, as well as the mean dissipated work for the frequency--dependent quantum harmonic oscillator. We use the formulas, 
\be
\la W\ra = \int dW\, W P(W) = -i \, G'(0) \ ,
\ee 
\be
\la W^2\ra = \int dW\, W^2 P(W) = - \, G''(0) \ ,
\ee 
\begin{equation}
\langle W_{dis}\rangle=\langle W\rangle-\Delta F\ .
\end{equation}
We consider the general expression (\ref{eq17}) for the characteristic function  of the work $G(\mu)$ as well as the various approximations (\ref{eq21}), (\ref{eq23}), (\ref{eq25}) and (\ref{eq28}) obtained in the high and low temperature limit and for adiabatic and nonadiabatic transformations. The free energy difference $\Delta F$ is given by Eq.~(\ref{18}). 
In the quantum limit $\hbar \beta \gg 1$, the free energy difference simplifies to 
\begin{equation}
\Delta F= \frac{1}{2}(\varepsilon_1-\varepsilon_0) \ ,
\end{equation}
whereas in the classical limit $\hbar \beta \ll 1$, one has
\begin{equation}
\Delta F=\frac{1}{\beta}\ln{\frac{\omega_1}{\omega_0}} \ .
\end{equation}
Note that for an  adiabatic transformation $Q^*=1$ and $\Delta\omega/\omega_0 \ll 1$, so that in this case $\Delta F \simeq \Delta\omega/ \beta$.
\begin{widetext}
\begin{tabular}{|c|c|c|c|}
\hline & & & \\  
 &$\langle W\rangle$&$\langle W^2\rangle-\langle W\rangle^2$&$\langle W_{dis}\rangle$ \\ 
& & & \\
\hline  General & & & \\
$Q^*\neq1$& $\frac{1}{2}(Q^*\varepsilon_1-\varepsilon_0)\coth{[\frac{\beta}{2}\varepsilon_0]}$ & $\frac{({Q^*}^2\varepsilon_1-\varepsilon_0)^2+({Q^*}^2-1)\varepsilon_1^2\cosh{[\beta\varepsilon_0]}}{4\sinh^2{[\frac{\beta}{2}\varepsilon_0]}}$ & $\frac{1}{\beta}\ln{\Bigg[\frac{(e^{\beta\varepsilon_0}+1)(\varepsilon_0-Q^*\varepsilon_1)+2(e^{\beta\varepsilon_0}-1)\frac{\sinh{[\frac{\beta}{2}\varepsilon_0]}}{\sinh{[\frac{\beta}{2}\varepsilon_1]}}}{2(e^{\beta\varepsilon_0}-1)}\Bigg]}$\\
Eq.~(\ref{eq17})& & &  \\

\hline  General & & & \\
$Q^*=1$& $\frac{1}{2}(\varepsilon_1-\varepsilon_0)\coth{[\frac{\beta}{2}\varepsilon_0]}$ & $\frac{1}{4}(\varepsilon_1-\varepsilon_0)^2\frac{1}{\sinh^2{[\frac{\beta}{2}\varepsilon_0]}}$ & $0$
\\Eq.~(\ref{eq17})& & &  \\

\hline $\hbar \beta\ll1$ & & & \\
$Q^*\neq1$ & $\frac{1}{\beta\omega_0}\big(Q^*\omega_1-\omega_0\big)$ & $\frac{1}{\beta^2\omega_0^2}\big(\omega_0^2-2Q^*\omega_0\omega_1+(2{Q^*}^2-1) \omega_1^2\big) $  & $\frac{1}{\beta}\big(\frac{Q^*\omega_1-\omega_0}{\omega_0}-\ln{[\frac{\omega_1}{\omega_0}]}\big)$\\
Eq.~(\ref{eq25}) & & & \\

\hline $\hbar \beta \ll1$ & & & \\
$Q^*=1$ &$\frac{1}{\beta\omega_0}\big(\omega_1-\omega_0\big)$ & $\frac{1}{\beta^2\omega_0^2}\big(\omega_1-\omega_0\big)^2$ & $0$ \\
Eq.~(\ref{eq21})& & & \\

\hline
$\hbar \beta\gg1$ & & & \\
$Q^*\neq1$& $\frac{1}{2}(Q^*\varepsilon_1-\varepsilon_0)$ & $\frac{1}{2}({Q^*}^2-1)\varepsilon_1^2$ & $\frac{1}{2}(Q^*-1)\varepsilon_1$\\
 Eq.~(\ref{eq28})& & & \\

\hline 
$\hbar \beta\gg1$ & & & \\
$Q^*=1$& $\frac{1}{2}(\varepsilon_1-\varepsilon_0)$ & $0$ & $0$\\
 Eq.~(\ref{eq23})& & & \\

\hline

\hline
\end{tabular}
\end{widetext}

\end{document}